\documentclass[9pt,twocolumn,twoside]{osajnl}

\journal{jocn} 

\setboolean{shortarticle}{false}

\usepackage{braket}
\title{Impact of topology on multipartite entanglement distribution protocols in quantum networks}

\author[1,2,*]{Jazz E. Z. Ooi}
\author[1]{Evan Sutcliffe}
\author[1,3]{Alejandra Beghelli}

\affil[1]{Optical Networks Group, University College London, London, WC1E 7JE, UK}
\affil[2]{Mathematical Institute, University of Oxford, Andrew Wiles Building, Woodstock Road, Oxford, OX2 6GG, UK}
\affil[3]{BT Group, Ipswich, IP5 3RE, UK}

\affil[*]{Corresponding author: ee.ooi.21@ucl.ac.uk}

\begin{abstract}

Quantum networks will rely on entanglement distribution to enable multi-user applications such as distributed quantum computing and cryptography.
While multipartite entanglement distribution routing protocols have been extensively studied on idealised grid topologies, less is understood about how real network structure shapes their performance and resource requirements.

We present a systematic study of four routing protocols for multipartite entanglement distribution, each characterised by the number of paths (single-path and multi-path) and routing strategy (star-based and tree-based), over 81 real network topologies.
We identified four distinct topology-dependent performance regimes, where: (i) all protocols perform poorly, (ii) tree-based protocols dominate, (iii) multi-path protocols dominate, or (iv) all protocols perform well.
By correlating clusters with graph metrics, we also provide structural explanations for the varied performance of specific protocols.

Additionally, motivated by the anticipated high cost of repeaters, we investigated the impact of repeater trimming on the performance of multi-path protocols. Topology strongly governs how far repeater nodes can be removed from the network while maintaining a given performance (distribution rate). 
For instance, in networks where only 80\% of nodes operate as repeaters, well-performing topologies are able to retain over 90\% of the distribution rate; whereas sparse, weakly connected graphs exhibit rapid performance degradation, retaining less than half of the distribution rate.

Our results provide a topology-aware framework for protocol selection and infrastructure optimisation in future quantum networks, bridging routing design with cost-aware deployment strategies.

\end{abstract}

\setboolean{displaycopyright}{false} 

\begin{document}

\maketitle

\section{Introduction}
\label{sec:intro}

The concept of a quantum network extends principles from classical communication theory, describing a collection of interconnected devices that can exchange quantum information \cite{Qnetwork}.
Communication of quantum information between two users can be achieved using shared entangled bipartite states, known as the Bell states \cite{bennett1996purification}. Although alternative communication strategies exist, entanglement-based schemes are favoured due to their tolerance against loss errors \cite{muralidharan2016optimal}. Using a shared Bell state, a qubit can be faithfully transmitted via a process known as quantum teleportation \cite{bennett1996purification}. More generally, multipartite entangled states enable communication among multiple users,
with applications in quantum key distribution \cite{appl_QKD}, secret ballots \cite{appl_secret}, quantum sensor networks \cite{appl_sensor} and distributed quantum computing \cite{appl_distrQC}.

A major challenge of quantum networking is that, when transmitted over long distances, photon absorption can cause the irreversible loss of information encoded in the (photonic) qubit \cite{max_range}.
Classically, this can be rectified by using amplifiers.
However, amplification or copying of qubit signals is impossible due to the no-cloning theorem \cite{no-cloning}, and thus direct transmission is infeasible.

A proposed method to overcome this problem is to use quantum repeaters. 
These act as intermediate nodes that extend communication range \cite{quantum_communications_book,azuma2023QuantumRepeater}, analogous to their classical counterparts (though with radically different working mechanisms).

There has been extensive research on bipartite cases to effectively distribute entanglement between two users in a repeater network for two-way quantum communication \cite{caleffi2017optimal,VanMeter2013Repeater,leone2024costvectoranalysis,Pirker2019stack,azuma2023QuantumRepeater,azuma2021tools,goodenough2021optimizing}. Work has also been done regarding multipartite entanglement routing on \emph{grid} topologies 
(also known as \emph{grid lattice} topologies)
\cite{IE3_evan_protocols,patil2021distance,natasha_thesis}. However, the symmetric node placement and uniform edge lengths of such networks make them unrepresentative of real-world networks, which often exhibit more complex topologies. Additionally, previous work on a limited number of real-world topologies has shown that different routing protocols for distributing multipartite states have varying effectiveness on different graph topologies \cite{IE3_evan_protocols,bugalho2023distributing}.

A further challenge is that current technology is insufficient regarding the production and usage of practical quantum repeaters, whose cost is anticipated to be high. This has led to interest in investigating the effect of repeater placement on entanglement distribution rates \cite{rabbie2022designing,avis2025optimization}. More generally, the problem of \emph{repeater resource allocation}, that is, the process of selecting optimal repeater locations over a quantum network, is a problem of interest to early quantum network design \cite{avis2025optimization}.

In this paper, we extend research on multipartite entanglement distribution protocols to investigate the impact of a diverse set of topologies based on real-world optical networks as well as repeater placement on the protocols' performance. For topology impact, we cluster  81 real-world topologies based on the performance of four chosen routing protocols, identifying connections between graph metrics and entanglement protocol performance. 
On the repeater placement aspect, we consider this problem with a focus on maximally entangled Greenberger-Horne-Zeilinger (GHZ) states due to applications in secret-sharing \cite{GHZ_secret} and quantum computing \cite{GHZ_QC}. Investigating a more realistic set of networks is motivated by the desire to reuse current classical network infrastructure when developing a future quantum internet \cite{rabbie2022designing}.

The rest of the paper is as follows. We first review previous work and relevant background in Section \ref{sec:litrev}.
We then describe our methodology in Section \ref{sec:method}.
Next, we go over the results obtained from identifying topology clusters and the impact of them on protocols' performance and repeater placement in Section \ref{sec:results}. 
Finally, we conclude our findings in Section \ref{sec:conclusion}.

\section{Literature Review}
\label{sec:litrev}
\subsection{Routing Protocols}
We study four multipartite entanglement distribution protocols. 
Each protocol 
can be specified by two variables: the number of paths explored and the routing strategy.
Combining the two variables gives us four possible protocols. These are single-path star-based (SPS), single-path tree-based (SPT), multi-path star-based (MPS), and multi-path tree-based (MPT) \cite{IE3_evan_protocols}.
All protocols assume \textit{timeslotted} operation as in Sutcliffe \textit{et al.} \cite{IE3_evan_protocols}, where for each timeslot, one attempt at Bell pair generation is performed per network edge. A successfully generated Bell pair, which is a maximally entangled state of two qubits stored in adjacent nodes, is termed an \textit{entanglement link}.

We consider protocols that distribute Bell pairs rather than generating a GHZ state in a node and sending the different qubits to the users. Using Bell pairs as a resource for long-range entanglement distribution can achieve higher distribution rates, as it allows for heralded generation schemes to attempt Bell pair generation over multiple timeslots \cite{azuma2023QuantumRepeater}. If multipartite state distribution was attempted directly, any photon loss would lead to failure, resulting in a much lower distribution rate. Furthermore, quantum networks designed around link-level Bell pair generation allow for greater operational flexibility \cite{Pirker2019stack}.

\subsubsection{Number of Paths: Single-path vs Multi-path}

\hspace{\parindent}\textit{Single-path (SP)}:
SP protocols work by precomputing a set of shortest paths
connecting all users. Entanglement link generation is only performed on edges within this set.
For multipartite cases, there are a few routing strategies (see Section 
\hyperref[subsubsec:route_strat]{\number\numexpr\value{section}\relax\ref*{subsubsec:route_strat}})
to precompute the routing solution (set of edges required to connect users). Entanglement link generation is then attempted on edges in the routing solution, and a 
GHZ state can be generated once Bell pairs are established between the users. 

\indent \textit{Multi-path (MP)}: 
MP protocols do not involve the precomputation of paths or trees before entanglement link generation. Instead, they attempt to generate a Bell pair over all network edges.
Assuming timeslotted entanglement link generation, the algorithm checks if the desired GHZ state can be established using only currently successfully generated entanglement links. Once the multipartite state has been distributed successfully, the protocol terminates.
The probabilistic nature of entanglement link generation means that each execution of an MP protocol could result in a different routing solution upon termination.
For the multipartite case, whether a given generated set of edges is `valid' depends on the routing strategy, as described in the following section.

\subsubsection{Routing Strategy: Star-based vs Tree-based}
\label{subsubsec:route_strat}
The \textit{routing strategy} determines when a valid routing solution exists, and thus the protocol can terminate. In this section and henceforth, we denote the set of users that wish to share a GHZ state as $S$ (and the number of users as $|S|$).

\textit{Star-based}:
Star-based (or \textit{Greedy}) protocols \cite{IE3_evan_protocols,bugalho2023distributing} 
use a fixed centre-node (see Section \hyperref[subsubsec:route_comp]{\number\numexpr\value{section}+1\relax\ref*{subsubsec:route_comp}} for precise definition and calculation algorithm), which may or may not be one of the users. These protocols only terminate once end-to-end entanglement is established over disjoint paths between each user and the centre-node. 
The disjointness requirement is based on the assumption that nodes can only store a single entanglement link per edge, which is consumed during the entanglement swapping task performed by quantum repeaters to achieve end-to-end entanglement. 
The union of the disjoint paths is a star graph (hence the name) with the users $S$ as leaves. 
By definition, star graphs only have a single node of degree greater than two, and thus require a single fusion operation to generate the GHZ state.

\textit{Tree-based}:
Tree-based (or \textit{Cooperative}) protocols \cite{IE3_evan_protocols,bugalho2023distributing} end if all users are connected through the minimum weight tree (known as the Steiner tree \cite{mehlhorn,kou}), where the edge weight is simply the edge length.
The Steiner tree requires up to $|S|-2$ fusion operations to generate a GHZ state \cite{IE3_evan_protocols}, compared to star-based protocols in which fusion is only performed at a single node. However, due to the fewer entanglement links used per GHZ state, tree-based protocols can achieve higher rates of entanglement distribution, assuming ideal local qubit operations, compared to their star-based counterpart \cite{IE3_evan_protocols}.

\begin{figure}[!htb]
\centering
\includegraphics[width=1.05\linewidth]{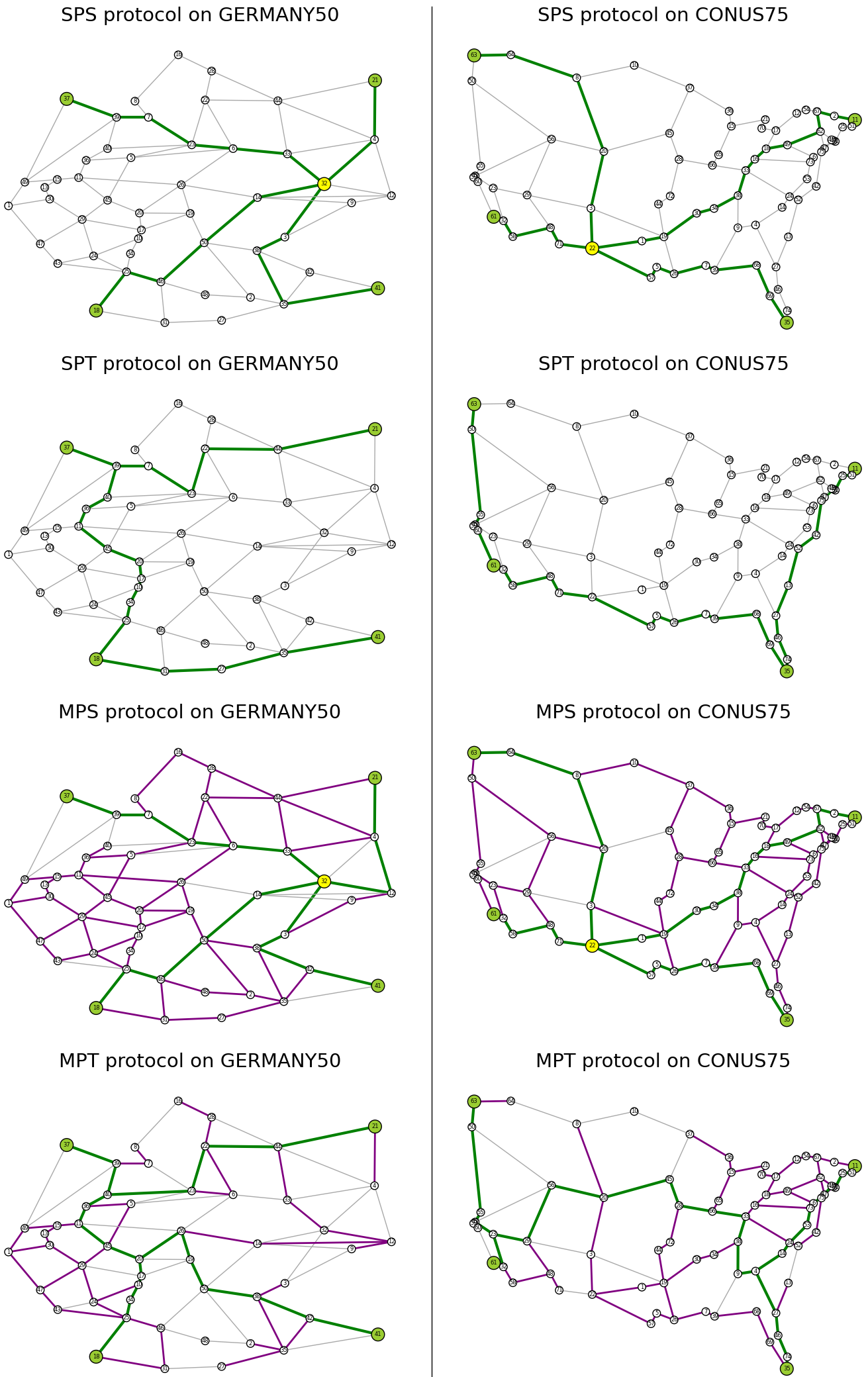}\vspace{+.5em}
\includegraphics[width=0.8\linewidth]{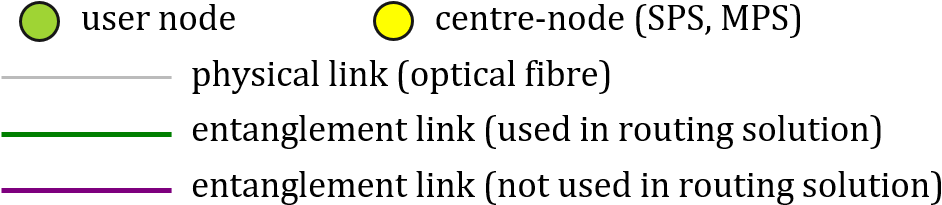}
\caption{Examples of routing solutions upon termination of each protocol on \textit{GERMANY50} and \textit{CONUS75}. The resulting routing solution is always identical for single-path protocols, but possibly different upon rerunning of multi-path protocols.}
\label{fig:protocol_ex}
\end{figure}

Figure \ref{fig:protocol_ex} illustrates routing solutions for the four multipartite entanglement distribution protocols, shown on the \textit{GERMANY50} and \textit{CONUS75} topologies.

\subsection{Multipartite Protocols Evaluation on Grid Topologies}

Previous work has shown that all MP routing protocols achieve better entanglement distribution rate than the SPS protocol in both bipartite and multipartite cases for grid topologies of all sizes and link generation probabilities \cite{IE3_evan_protocols,natasha_thesis}.
At low link generation probabilities, star-based SPS and MPS protocols perform similarly.

However, due to the route precomputation involved with SP protocols, they require fewer quantum repeaters to distribute multipartite states. This may be useful when considering that repeater costs are expected to be high. 
In larger topologies, an increased proportion of repeaters are minimally used or unused altogether if, as expected in the early stages of deploying quantum networks, only a subset of nodes request entanglement \cite{rabbie2022designing}. This led to studies on the removal of repeaters from the network according to their usage rate, starting from the least used repeaters \cite{natasha_thesis}. 

In Siow \textit{et al.}, it was shown that MPS converges to SPS (labelled `SP' in the cited paper) as more repeaters are removed, since the most commonly used repeaters lie on the shortest paths used by SPS \cite{natasha_thesis}.
On the other hand, MPT allows the removal of more repeaters whilst retaining better performance than both SPS and MPS (SPT was not considered). For both MP protocols, for a given performance, a larger grid leads to a higher number of removable repeaters.


Repeater removal can be applied by instructing the routing protocol to disregard certain repeaters, saving resources by generating entanglement links over a smaller set of edges. This approach allows for a more efficient use of network resources.

There has been minimal extension of this work to more realistic topologies. For instance, this is done in Sutcliffe \textit{et al.}, albeit for a small set of six real-world topologies, without investigation of repeater removal \cite{IE3_evan_protocols}.
We seek to expand these findings to a larger set of 81 real-world topologies, as well as applying the results to optimise repeater allocation.

\section{Methodology}
\label{sec:method}
\subsection{Network Model}
A quantum network can be represented as an undirected graph $G=(V,E)$, with $V$ and $E$ denoting the set of vertices and edges in $G$ respectively. 

Each node (or vertex) $v\in V$ is a quantum device, equipped with a quantum memory associated with each connected edge.
Our work assumes that qubits remain coherent while stored in quantum memories. This will be feasible for logically protected qubits, or if the decoherence time is large in comparison to the time required to distribute a GHZ state. As such, we assume that after successful generation, entanglement links remain stored until the protocol successfully distributes the multipartite state. We also assume that each node is able to perform local operations and classical communication (LOCC), as well as act as a repeater. Nodes can act as both a user and a repeater simultaneously. Repeater functionality includes entanglement link generation, entanglement swapping and entanglement fusion \cite{azuma2023QuantumRepeater}, described in Section 
\hyperref[sec:QOp]{\number\numexpr\value{section}\relax\ref*{sec:QOp}}.

Each edge $e\in E$ represents a physical link (an optical fibre cable) connecting two nodes over which qubits can be transmitted. 
The vertices in a subset $S\subseteq V$ are defined as \textit{users}, who request a multipartite entanglement state to be distributed between them to enable the exchange of quantum information. Throughout, we consider distribution to be attempted between a fixed set of $|S|=4$ users in each network topology (for comparison with previous work on grid topologies \cite{IE3_evan_protocols,patil2021distance,natasha_thesis}, since the number of users is bounded by the maximum node degree).

\subsection{Quantum Operations}\label{sec:QOp}
\textit{Entanglement link generation} is performed between adjacent nodes over an edge. Entanglement links are assumed to be generated in the form of a Bell pair: 

\begin{equation}
\ket{\Phi^+} = \frac{1}{\sqrt{2}}(\ket{00} + \ket{11})
\label{eqn:bell}
\end{equation}

Generation of entanglement links is probabilistic. The probability of success $p_e$ is modelled as:
\begin{equation}
p_e(L) = p_{\text{tr}}(L) \times p_{\text{op}}
\label{eqn:p_e}
\end{equation}

where $p_{\text{tr}}(L) = 1-p_{\text{loss}}(L)$ is the \textit{transmission} probability, with $p_{\text{loss}}$ denoting the probability of qubit loss during transmission. Assuming photonic qubits transmitted over optical fibres of length $L$ km, and attenuation of $0.2$ dB km$^{-1}$ \cite{kanamori2021transmission}: 
\begin{equation}
p_{\text{tr}}(L) = 10^{-0.2L/10} = \exp\left(-\frac{0.2\ln10}{10}L\right)
\label{eqn:p_tr}
\end{equation}

$p_{\text{op}}$ is the probability associated with imperfect operations during entanglement generation. We disregard this in our simulation, taking $p_{\text{op}}=1$.

\begin{figure}[htbp]
\centering
\includegraphics[width=0.7\linewidth]{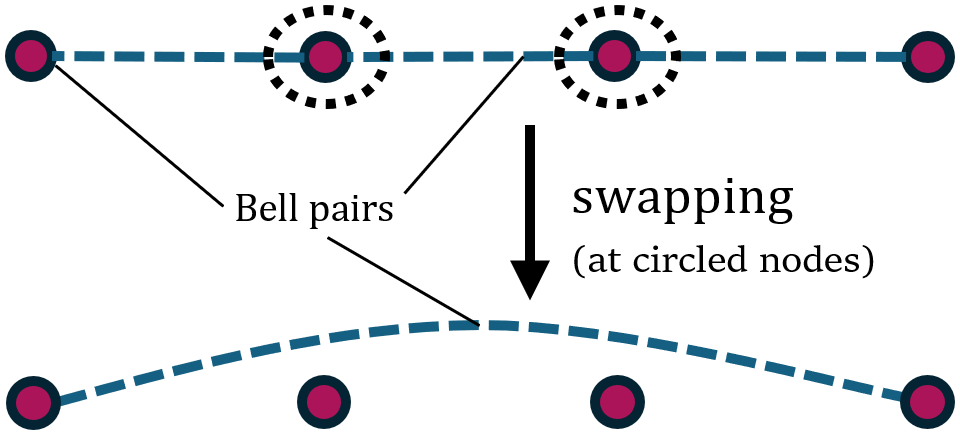}
\caption{Entanglement swapping used to establish a long-distance entanglement link by performing BSMs on the qubits held at circled nodes.}
\label{fig:swapping}
\end{figure}

\textit{Entanglement swapping} can then be used to `combine' adjacent entanglement links, to form entanglement links shared between non-adjacent nodes.
In a path where every edge has an entanglement link, as illustrated in Figure \ref{fig:swapping}, performing Bell state measurements (BSMs) on intermediate nodes allows the distribution of Bell pairs between the two longer-distance end-points \cite{azuma2023QuantumRepeater}. 
Assuming ideal swapping, it is more rate-efficient to generate shorter sublinks then perform BSMs, than to attempt entanglement link generation directly over a long distance. This is not necessarily true in the non-ideal case.

Just as a Bell pair is a maximally entangled 2-qubit state, a Greenberger-Horne-Zeilinger (GHZ) state is a maximally entangled $N$-qubit state \cite{gisin1998bell}. 


\begin{equation}
|\text{GHZ}_N\rangle = \frac{1}{\sqrt{2}}\left(|0\rangle^{\otimes N} + |1\rangle^{\otimes N}\right)
\label{eqn:ghz}
\end{equation}

\textit{Entanglement fusion} takes two GHZ states of size $n_1$ and $n_2$, entangling a qubit from each state by measuring one of them and performing corrections according to the measurement result.
Repeated iterations allow the creation of $N$-qubit GHZ states by combining $N-1$ Bell pairs ($N>2$). This follows the definition in \cite{DeBone2020GHZStates}.

\begin{figure*}[htb]
\centering
\includegraphics[width=0.72\linewidth]{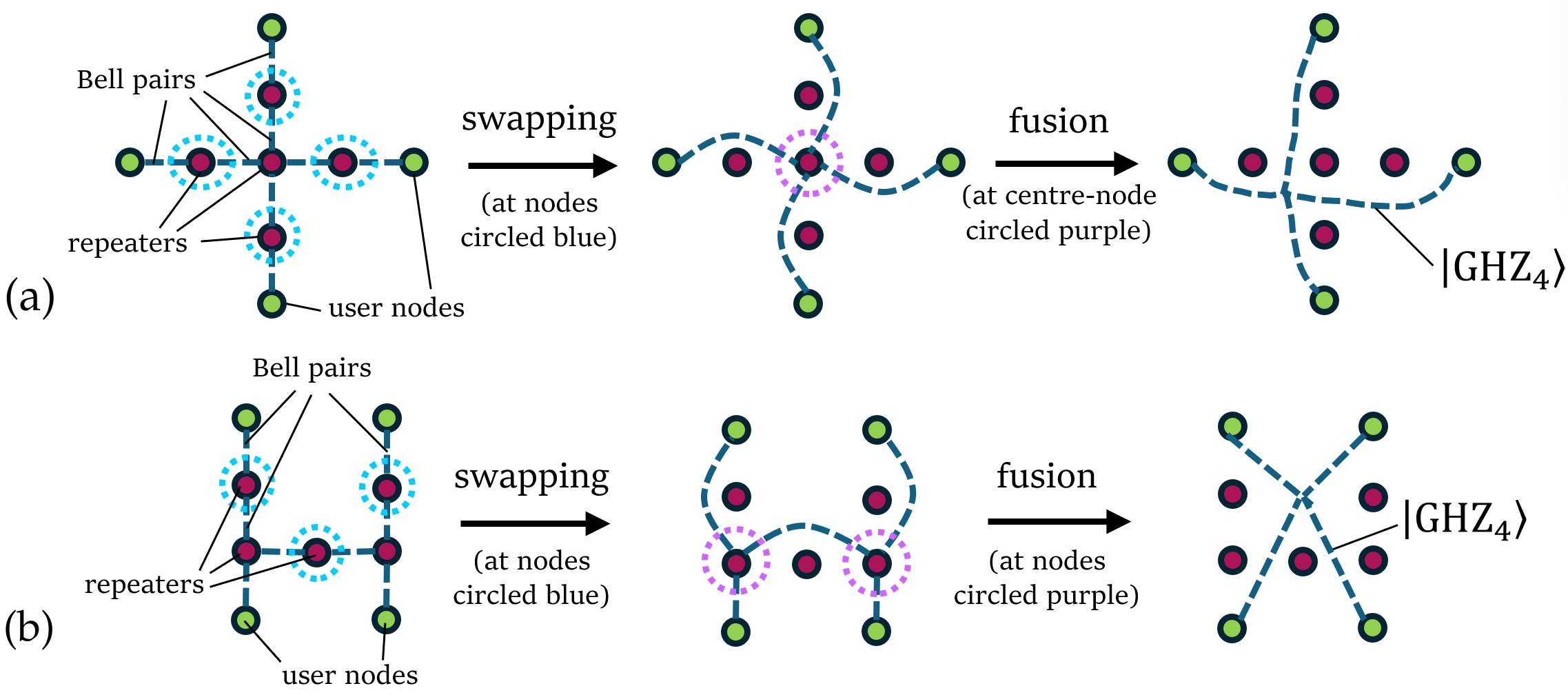}
\caption{Illustration of entanglement swapping and fusion to combine 8 Bell pairs into a $|\text{GHZ}_4\rangle$ state between user nodes, (a) with a single fusion operation at a centre-node, typical of star-based protocols; and (b) multiple fusion operations, typical of tree-based protocols. Blue dashed lines between two nodes represent shared Bell pairs, and a star-shaped configuration with $N$ arms represents entanglement between $N$ qubits in a $\ket{\text{GHZ}_N}$ state.}
\label{fig:fusion}
\end{figure*}

Entanglement swapping and fusion can be used to combine multiple shorter Bell pairs into a single GHZ state, as illustrated in Figure \ref{fig:fusion}. 

Each local quantum operation required can be modelled as having a negative impact on the rate or fidelity of distributed entangled states \cite{azuma2021tools,patil2021distance}. In current quantum networks,
qubit operations remain imperfect and resource-intensive \cite{muralidharan2016optimal}. As such, approaches that minimise the number of required operations is of importance for feasible near-term implementation.
However, since our work focuses on the generation of links that make up the GHZ state, we assume Bell pairs are logically protected, and that there is no cost associated with local quantum operations.

\subsection{Multipartite Entanglement Distribution Protocols} \label{sec:protocols}

The multipartite entanglement distribution problem can be formalised as follows. Given a graph $G=(V,E)$ with a subset of user nodes $S\subseteq V$, where $|S|=N\geq 3$, generate a $\ket{\text{GHZ}_N}$ state between all users, maximising the rate of generation.

Such a state can be created using LOCC operations, provided a routing solution (consisting of successfully generated entanglement links) that connects all users exists.

\subsubsection{Routing Solution Computation}
\label{subsubsec:route_comp}
For star-based protocols, we use a max-flow routing algorithm \cite{dinic} to identify the shortest disjoint paths from a candidate centre-node to all users. 
We calculate the choice of centre-node as follows: for each valid candidate centre-node $v\in V$, we connect a super-source to the centre-node and a super-sink to each user, and choose as the centre-node the node that maximises flow whilst minimising cost (edge lengths). For the SP protocols, this route is selected once, at the start of the protocol. For the MP protocols, routing is reattempted at the end of every generation timeslot, over edges that hold Bell pairs. 

Tree-based protocols simply require all user nodes $u\in S$ to be connected, implying the existence of a tree connecting them. The least-cost tree (with respect to total edge length) is known as the \textit{Steiner tree}. Solving the Steiner tree problem is NP-hard \cite{steiner_NP}, and so we use the Kou \cite{kou} and Mehlhorn \cite{mehlhorn} approximate Steiner tree algorithms. 
As heuristics, each approach may yield different solutions, with no consistent winner; as such, we run both algorithms, selecting the lower cost (with respect to the sum of edge lengths) tree as the routing solution.

\subsubsection{Quantifying Protocol Performance}
To evaluate protocol performance, Monte Carlo simulations are run for each protocol on each of the 81 topologies selected. A total of 5000 GHZ states are generated for each protocol-topology pair, 
with each simulation iteration attempting entanglement link generation for multiple timeslots. Once an entanglement link is generated successfully, it is stored until a GHZ state can be distributed and the protocol terminates.

Each protocol, upon successful termination, returns a valid routing solution that can be used to generate a GHZ state, as well as the number of timeslots until generation success or termination. Performance of a protocol on a given graph $G=(V, E)$
is quantified by the \emph{expected number of timeslots} $\mathbb{E}[T]$ needed to generate a successful routing solution (the expected waiting time); as well as the \emph{distribution rate} $\lambda= 1/\mathbb{E}[T]$, i.e.\ how many valid GHZ states are generated on average per timeslot. $T$ is a random variable corresponding to the time required to distribute a single GHZ state.





\subsection{Dataset of Real Networks}
The investigation of existing real-world optical networks is motivated by the fact that a cost-effective way of deploying quantum networks uses existing classical infrastructure, using deployed dark fibre and installing quantum repeaters at selected locations \cite{rabbie2022designing}.

The comprehensive set of real-world optical networks we work with was sourced from the \textit{Topology Bench} dataset, which combines previously published datasets from 
various papers and repositories \cite{topobench_dataset}.
Each edge corresponds to a physical optical fibre link, and each node corresponds to an exchange.

Out of the 105 real optical networks provided by \textit{Topology Bench}, 81 topologies were selected for analysis. We discarded 24 networks due to their unsuitability for GHZ state distribution protocols or analysis. These include ring networks, which have no valid centre-node since all nodes have degree two; star-like graphs, where all protocols yield the same result; and barbell-like graphs \cite{barbell}, which have no valid centre-node for more than two users on either side (since paths must be disjoint).  

The exponential decay of entanglement link success probability with respect to link length (\eqref{eqn:p_tr}) implies impractical simulation times for long edges. For example, in the \textit{JPN25} network, the edge between Sapporo and Niigata is 1396.65 km long, corresponding to a success probability of $p_e(1396.65)=1.17\times10^{-28}$. Due to this impractically low probability, we scaled down topologies in a way such that the longest edge in the network is 100 km, which bounds entanglement link generation probability by $p_e(100)= 0.01$.
This rescaling implies that our results reflect network topology structure rather than absolute physical scale.

\subsection{\textit{k}-Means Clustering}
To identify differences in protocol performance across networks, 
we employed the $k$-means \cite{macqueen1967multivariate} unsupervised machine learning technique to cluster the 81 topologies, using performance (namely $\mathbb{E}[T]$) of the four protocols as training features.
To determine the optimal number of clusters $k$, multiple methods were used, including the elbow \cite{thorndike1953elbow} and silhouette \cite{rousseeuw1987silhouettes} methods.
We found that $k=4$ was the best value in the context of our study.

\subsection{Repeater Trimming}
To identify the minimum number of repeaters and their locations for a given performance of MP protocols, we simulated repeater \textit{trimming}. Under trimming, repeater functionality is iteratively removed from nodes with the least (possibly never) used repeaters.

We refer to repeaters that are never utilised in a routing solution as \textit{redundant} repeaters (or nodes), and the remaining repeaters as \textit{active} repeaters. Note that redundant and active repeaters can differ for each protocol.

For SP protocols, any repeater that is not part of the precomputed routing solution is redundant. Hence, trimming simulation for SP protocols does not offer any insight; there is one iteration where all unused repeaters are trimmed (which does not affect performance), after which no further trimming can be done.

In contrast, MP protocols dynamically select repeaters across multiple candidate paths and thus utilise a broader range of repeater nodes throughout the network. In this case, redundant repeaters may arise from branches extending away from primary routing paths (e.g.\ legs of star-like topologies with more than four arms) or from nodes located along paths disproportionately longer than alternatives.

After initially trimming all redundant nodes, we run 5000 Monte Carlo simulations on the newly trimmed topology. Each subsequent iteration sees trimming of the least used active repeaters, then rerunning of 5000 simulations. This process is repeated until the topology is maximally trimmed, meaning all remaining repeaters are always in use.

We evaluate the impact of repeater trimming by considering how many active repeaters can be trimmed while retaining a fraction of the distribution rate of the untrimmed graph.
We denote this fraction by $\lambda/\lambda_0$ , where $\lambda$ and $\lambda_0$ (for each topology) denote the distribution rate on the trimmed and untrimmed graphs, respectively. For instance, a value of $\lambda/\lambda_0=0.6$ indicates that the graph is trimmed until any further trimming causes $\lambda$ to fall below 60\% of $\lambda_0$.

\begin{figure}[ht]
\centering
\includegraphics[width=0.9\linewidth]{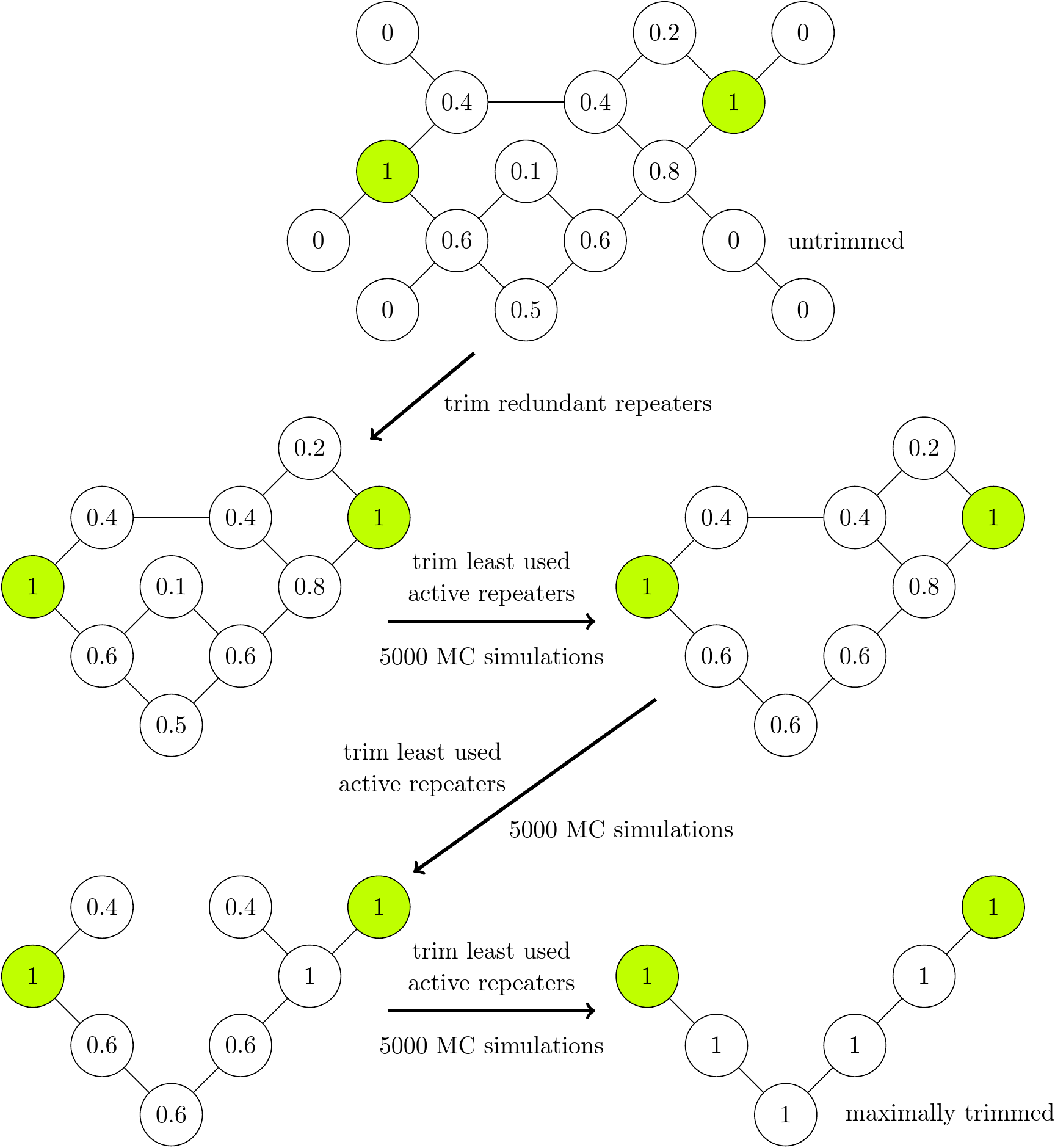}
\caption{Example of the repeater trimming process on a simple graph. User nodes are coloured green. The number (between 0 and 1) on each node is the fraction of Monte Carlo runs (out of 5000) that utilises the repeater.}
\label{fig:trimming_ex}
\end{figure}

\begin{figure}[ht]
\centering
\includegraphics[width=0.99\linewidth]{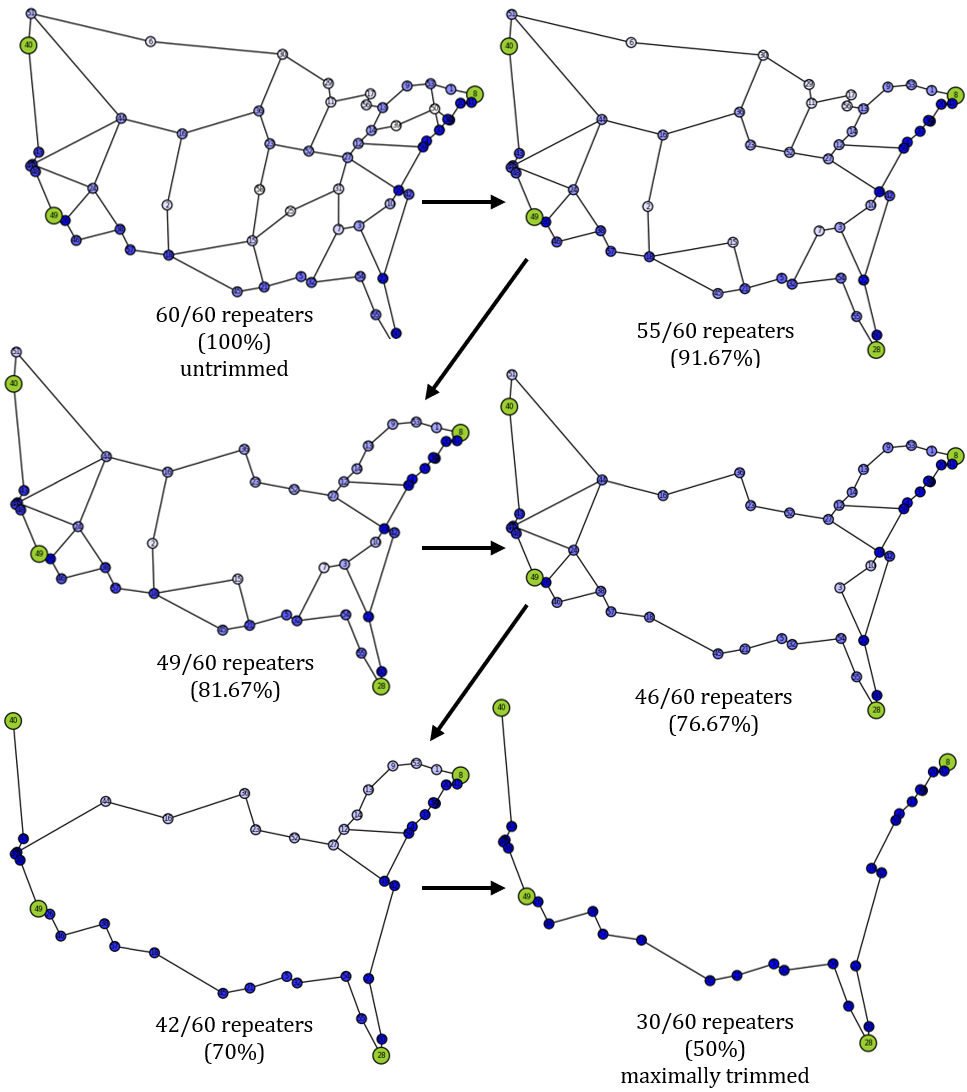}
\caption{Repeater trimming process for MPT on topology \textit{CORONET} (Cluster 4). Shade of blue corresponds to repeater usage for a given (possibly trimmed) topology. Darker blue indicates higher usage, and unused repeaters are white. Each trimming step (arrow) involves trimming of the least-used active repeaters, and 5000 Monte Carlo simulations on the newly trimmed topology. This topology has no redundant repeaters.}
\label{fig:trimming_ex2}
\end{figure}

Figures \ref{fig:trimming_ex} and \ref{fig:trimming_ex2} illustrate the trimming process for a simple topology and a real topology (\textit{CORONET}) respectively.

\section{Results}
\label{sec:results}

\subsection{Protocol Evaluation}
Protocol performance is quantified by average waiting time (timeslots until termination) $\mathbb{E}[T]$ required to distribute GHZ state, as well as its reciprocal, the distribution rate $\lambda$.

We define the set of users as the four nodes furthest apart from each other (maximising the sum of pairwise shortest path distances).
See, for example, the green user nodes in Figure \ref{fig:protocol_ex}.



\begin{figure}[H]
\centering
\includegraphics[width=0.7\linewidth]{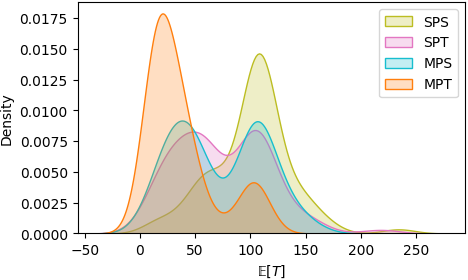}
\caption{Kernel Density Estimation (KDE) plots of $\mathbb{E}[T]$ for the studied protocols, on all topologies.}
\label{fig:kde}
\end{figure}

Performance of each protocol is shown in Table \ref{tab:cluster_perf}, and the distribution of $\mathbb{E}[T]$ is shown in Figure \ref{fig:kde}. 
On average, MPT performs best (mean $\mathbb{E}[T]=37.25$), and SPS performs worst (mean $\mathbb{E}[T]=100.84$). Given the same routing strategy (star-based or tree-based), MP always outperforms SP, since SP restricts the search space to the single precomputed optimal solution, whereas MP considers all solutions (including the optimal one). Given equal resources, tree-based protocols always outperform the star-based variant, since any star-based routing solution also satisfies the tree-based requirement (but not vice versa). SPT and MPS perform similarly overall (mean $\mathbb{E}[T]$ of 75.09 and 72.39 respectively), and when exactly they outperform each other is dependent on the specific topology.


\begin{table}[H]
\centering
\caption{\bf $\mathbb{E}[T]$ (number of slots) per cluster and protocol}
\begin{tabular}{crrrr}
\hline
 & \multicolumn{1}{c}{SPS} & \multicolumn{1}{c}{SPT} & \multicolumn{1}{c}{MPS} & \multicolumn{1}{c}{MPT} \\ \hline
Cluster 1 & 129.49$^\dagger$ & 122.34$^\dagger$ & 116.96$^\dagger$ & 90.32$^\dagger$ \\ 
Cluster 2 & 114.77$^\dagger$ & 35.18$^*$ & 106.90$^\dagger$ & 21.82$^*$ \\ 
Cluster 3 & 109.56$^\dagger$ & 95.37$^\dagger$ & 68.91$^*$ & 28.76$^*$ \\ 
Cluster 4 & 70.00$^*$ & 46.13$^*$ & 33.52$^*$ & 18.92$^*$ \\
\hline
Overall & 100.84\phantom{$^\dagger$}  & 75.09\phantom{$^\dagger$}  & 72.39\phantom{$^\dagger$}  & 37.25\phantom{$^\dagger$} \\
\hline
\end{tabular}
\begin{flushleft}
\footnotesize 
\quad *: Fast generation and good performance (better than overall protocol average)\\
\quad $^\dagger$: Slow generation and poor performance (worse than overall protocol average)
\end{flushleft}
\label{tab:cluster_perf}
\end{table}

\vspace{-2em}

\subsection{Topology Clustering}
A clustering approach was used to characterise differences in protocol performance across various topologies and to examine its relationship with graph metrics such as size, density, and eccentricity. Table \ref{tab:clusters} lists the topologies belonging to each cluster. 


\begin{figure}[H]
\centering
\includegraphics[width=0.99\linewidth]{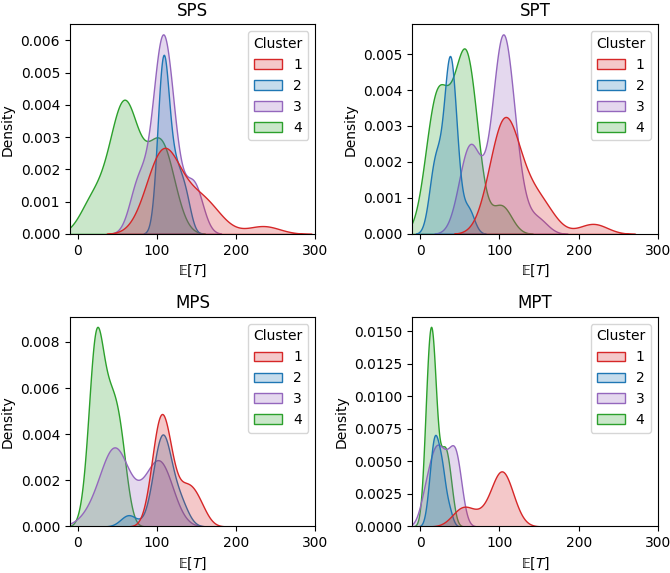}
\caption{KDE plots of $\mathbb{E}[T]$ for various clusters and protocols. Recall that a protocol performing `well' corresponds to a lower expected waiting time $\mathbb{E}[T]$.}
\label{fig:kde_clusters}
\end{figure}

\begin{figure}[H]
\centering
\includegraphics[width=0.99\linewidth]{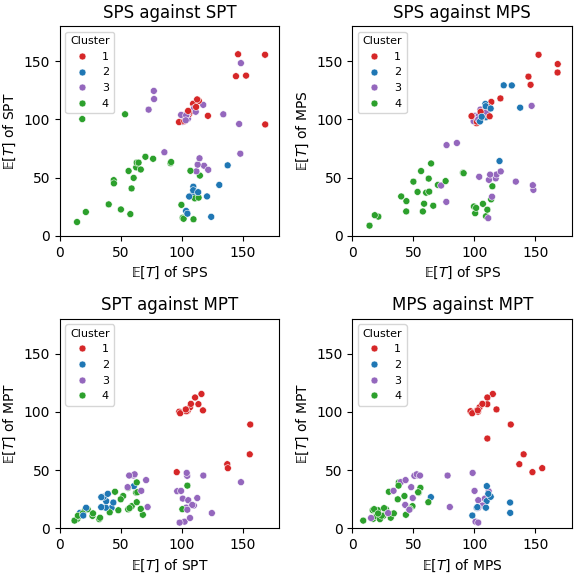}
\caption{Scatter plots illustrating topology clustering with respect to $\mathbb{E}[T]$.}
\label{fig:scatter_clusters}
\end{figure}

\begin{table}[H]
\centering
\caption{\bf Topologies classified into each cluster}
\begin{tabular}{p{2cm}>{\raggedright\arraybackslash}p{5.75cm}}
\hline
\textbf{Cluster 1} \newline (17 topologies) &
{\small TATANID, GTSCZECHREPUBLIC, GTSPOLAND, RENATER1999, ANS, BICS, HIBERNIAGLOBAL, LONI, BIZNET, JGN2PLUS, JANOSUS, MZIMA, ISTAR, CONUS30, GAMBIA, CERNET, GBLNET}
\\
\hline
\textbf{Cluster 2} \newline (12 topologies) & 
{\small ERNET, JPN12, ARNES, VIA, REDIRIS, OXFORD, NETWORKUSA, IBM, PACKETEXCHANGE, PIONIER, GETNET, PIONIERL3 }\\
\hline
\textbf{Cluster 3} \newline (24 topologies) & 
{\small SANET, LAYER42, DT17, CANARIE, RENATER2001, CWIX, RENATER2008, PORTUGAL, DIGEX, ITALY, REDCLARA, RENATER2006, INTEGRA, COST, CRLNETWORKSERVICES, CLARANET, EON, RENATER2010, CONUS6077, METRONA, JPN25, ELIBACKBONE, POLSKA, NOBELEU}
 \\
\hline
\textbf{Cluster 4} \newline (28 topologies) & 
{\small BEYONDTHENETWORK, NSFNET13, CANARIE, RENATER2004, CESNET, GERMANY50, NETRAIL, GEANT2, RENATER, CONUS6079, CONUS75, USA100, JPN48, CONUS100, ARPANET, PALMETTO, IRIS, MEMOREX, HIBERNIAUS, KAREN, NOBELGERMANY, NOBELUS, SPIRALIGHT, GRNET, RAILTELINDIA, CORONET, GEANT, NOEL}
 \\
\hline
\end{tabular}
\label{tab:clusters}
\end{table}

\newpage

As described in Table \ref{tab:cluster_perf} and illustrated in Figures \ref{fig:kde_clusters}--\ref{fig:scatter_clusters},
each cluster label can be interpreted as follows:
\begin{itemize}[itemsep=0pt]
    \item Cluster 1: \textit{Globally Adverse Topologies}.\\
    All protocols perform poorly.
    \item Cluster 2: \textit{Tree Dominant Topologies}.\\
    Tree-based protocols perform well.\\
    Star-based protocols perform poorly.
    \item Cluster 3: \textit{Multi-path Dominant Topologies}.\\
    Multi-path protocols perform well.\\
    Single-path protocols perform poorly.
    \item Cluster 4: \textit{Globally Favourable Topologies}.\\
    All protocols perform well.
\end{itemize}

Here, topologies classified as `poor' or `good' performing have $\mathbb{E}[T]$ greater than or less than the overall mean, respectively.

Various network metrics averaged globally and across each cluster can be seen in Table \ref{tab:cluster_metrics}.
The Python package NetworkX \cite{networkx} was used for simulation and computation of all metrics. 

\begin{table}[H]
\centering 
\caption{\bf Mean values of various graph attributes and metrics, overall and for each cluster}
\setlength{\tabcolsep}{2.5pt}
\resizebox{0.999\linewidth}{!}{
\begin{tabular}{lrrrrr}
\hline
\multicolumn{1}{l}{\textbf{Cluster}} & \multicolumn{1}{c}{All} & \multicolumn{1}{c}{1} & \multicolumn{1}{c}{2} & \multicolumn{1}{c}{3} & \multicolumn{1}{c}{4}\\
\hline\hline
\multicolumn{6}{l}{\bf Graph size:}\\
Number of Nodes & 29.136 & 31.647 & 18.333 & 25.917 & 35.000 \\
Number of Edges & 39.309 & 39.294 & 23.083 & 35.042 & 49.929 \\
\hline
\multicolumn{6}{l}{\bf Edge lengths \& Eccentricity (km):}\\
Mean Edge Length & 35.203 & 36.833 & 36.191 & 34.513 & 34.381 \\
Mean Distance & 123.382 & 156.661 & 106.419 & 111.594 & 120.551 \\
Diameter & 299.990 & 387.883 & 256.071 & 272.892 & 288.676 \\
Radius & 172.271 & 219.422 & 150.879 & 157.923 & 165.110 \\
Edge Length Std.\ Dev.\ & 22.196 & 22.336 & 23.723 & 20.955 & 21.874 \\
\hline
\multicolumn{6}{l}{\bf Centrality \& Density:}\\
Density & 0.137 & 0.115 & 0.181 & 0.142 & 0.128 \\
Global Efficiency & 0.403 & 0.368 & 0.458 & 0.413 & 0.391 \\
Mean Laplacian Centrality & 0.120 & 0.115 & 0.158 & 0.119 & 0.107 \\
Mean Closeness Centrality & 0.320 & 0.283 & 0.371 & 0.331 & 0.312 \\
\hline
\multicolumn{6}{l}{\bf Connectivity:}\\
Node Connectivity & 1.383 & 1.176 & 1.250 & 1.458 & 1.500 \\
Mean Node Degree & 2.655 & 2.366 & 2.548 & 2.740 & 2.803 \\
\hline
\end{tabular}
}
\label{tab:cluster_metrics}
\end{table}

Recall that $V$ and $E$ denote the set of vertices and edges respectively of graph $G=(V,E)$.
Distance $d(v,u)$ between two nodes $v,u\in V$ is the length of the shortest path between them \cite{graphtheory_book}. 

Eccentricity of node $v\in V$ is the greatest distance between $v$ and any other node in $G$.
\begin{equation}
    \text{eccentricity}(v) = \max_{u\in V} d(v,u)
\end{equation}

Diameter and radius of $G$ are the maximum and minimum eccentricity respectively \cite{graphtheory_book}. Diameter is thus also the greatest distance between any pair of nodes in $G$.
\begin{align}
    \text{diameter}(G) &= \max_{v\in V}\max_{u\in V} d(v,u)\\
    \text{radius}(G) &= \min_{v\in V}\max_{u\in V} d(v,u)
\end{align}

The density of a graph is defined as $|E|$ over the maximum possible number of edges $\binom{|V|}{2}$ \cite{graphtheory_book}.
\begin{equation}
    \text{density}(G) 
    = \frac{|E|}{\binom{|V|}{2}}
    = \frac{2|E|}{|V|\left(|V|-1\right)}
\end{equation}

The efficiency of two nodes is the reciprocal of their distance.
Global efficiency of a graph is the average efficiency between all such pairs, which quantifies the efficiency of a network in exchanging information \cite{global_eff}.
\begin{equation}
    \text{global efficiency}(G) 
    = \frac{1}{|V|\left(|V|-1\right)}
    \sum_{\substack{v,u\in V \\ v\neq u}}\frac{1}{d(v,u)}
\end{equation}

Laplacian centrality of $v\in V$ is the change in the sum of squared eigenvalues of the graph's Laplacian matrix if $v$ is removed \cite{laplacian_centrality}, quantifying a node's structural importance to the network.
Closeness centrality of $v\in V$ is the reciprocal of its average distance to all other nodes \cite{centrality}, quantifying how close the node is to the rest of the network.

\begin{equation}
    \text{closeness centrality}(v) = \left(
    \frac{1}{|V|-1}\sum_{\substack{u\in V \\ u\neq v}}d(v,u)
    \right)^{-1}
\end{equation}

Finally, node connectivity is the minimum number of nodes removed to disconnect a graph \cite{vertex_connectivity}, quantifying network resilience to node failures.

Based on these metrics, we get insight into the type of topologies falling into each cluster.

\begin{figure}[H]
\centering
\includegraphics[width=0.495\linewidth]{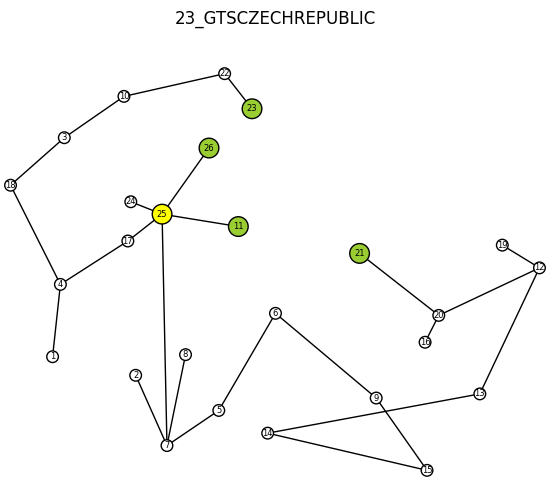}
\includegraphics[width=0.495\linewidth]{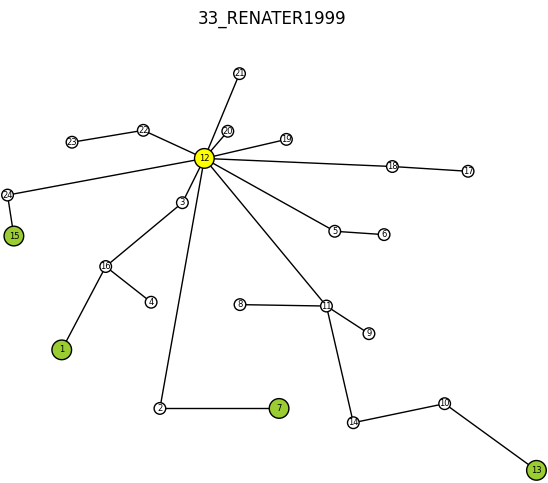}
\caption{\textit{GTSCZECHREPUBLIC} and \textit{RENATER1999}, two topologies in Cluster 1.}
\label{fig:cluster1}
\end{figure}

Topologies in Cluster 1 (\textit{globally adverse topologies}) have long average edge lengths that contribute to high eccentricity values 
(mean diameter and radius 29.30\% and 27.37\% higher than global mean respectively) and very low
density and connectivity (16.06\% and 14.97\% lower than global mean). For instance, the long path between the leftmost and rightmost users for \textit{GTSCZECHREPUBLIC} and \textit{RENATER1999} topologies, shown in Figure \ref{fig:cluster1}, bottlenecks performance for all protocols.

\begin{figure}[H]
\centering
\includegraphics[width=0.495\linewidth]{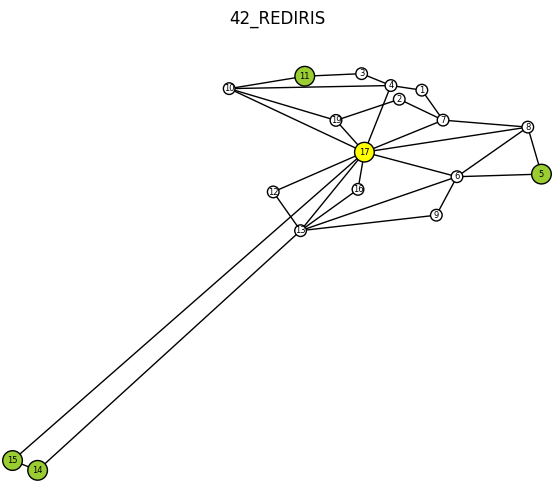}
\includegraphics[width=0.495\linewidth]{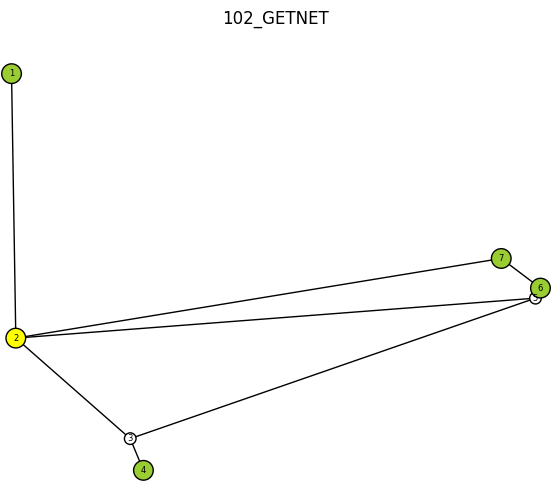}
\caption{\textit{REDIRIS} and \textit{GETNET}, two topologies in Cluster 2. Note their relatively small size and sparse edge length distribution.}
\label{fig:cluster2}
\end{figure}

Cluster 2 (\textit{tree dominant topologies}) consists of dense, small topologies (mean density 32.12\% higher, number of nodes and edges 37.08\% and 41.28\% lower than global mean), on which tree-based protocols dominate.
Edge lengths are sparsely distributed (standard deviation 6.88\% higher than global mean).
Typically, these topologies have multiple users in a region of the graph that is separated from the denser part of the graph (which contains the centre-node and the other users) by long edges, and hence long paths. For example, referring to Figure \ref{fig:cluster2}, these are the two bottom-left users in \textit{REDIRIS}, and the two rightmost users in \textit{GETNET}.
The disjoint constraint forces central-node dependent star-based protocols to suboptimally use one long edge per user with low entanglement link generation probability, leading to poor performance. In contrast, tree-based protocols require only a single long edge, since Steiner trees only require a single edge to connect the users back to the denser region.

\begin{figure}[H]
\centering
\includegraphics[width=0.495\linewidth]{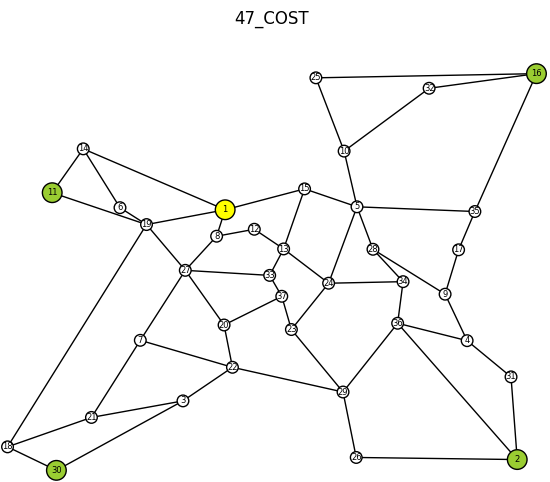}
\includegraphics[width=0.495\linewidth]{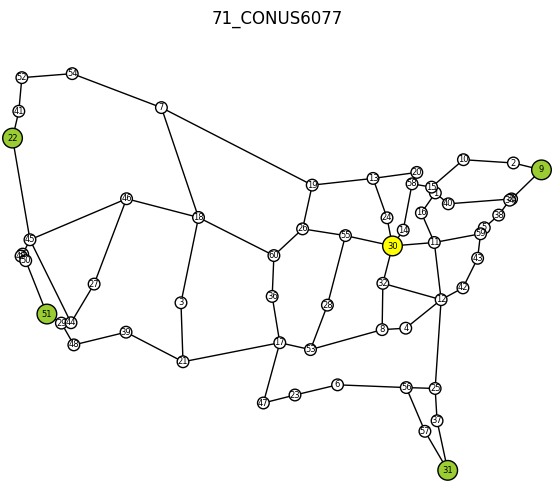}
\caption{\textit{COST} and \textit{CONUS6077}, two topologies in Cluster 3.}
\label{fig:cluster3}
\end{figure}

Cluster 3 (\textit{multi-path dominant topologies}) has topologies with more uniform edge lengths (standard deviation 5.59\% lower than global mean), showing similarities to grid-like topologies.
Multi-path protocols dominate due to the existence of multiple routing solutions of similar total cost. The single-path restriction, where generation is only attempted on a single precomputed route, leads to suboptimal performance. 
For example, note the relatively uniform edge lengths of \textit{COST} and \textit{CONUS6077}, as seen in Figure \ref{fig:cluster3}.

\begin{figure}[H]
\centering
\includegraphics[width=0.495\linewidth]{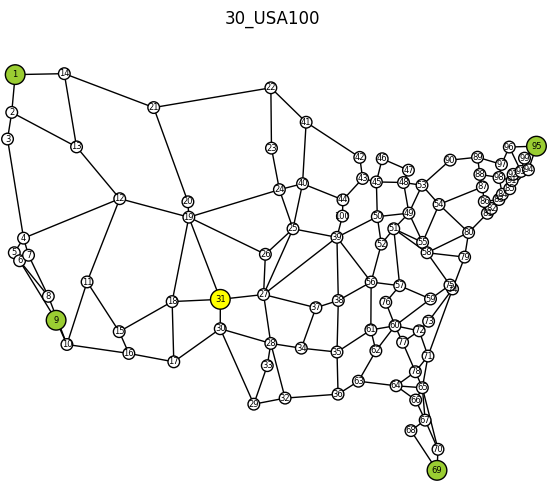}
\includegraphics[width=0.495\linewidth]{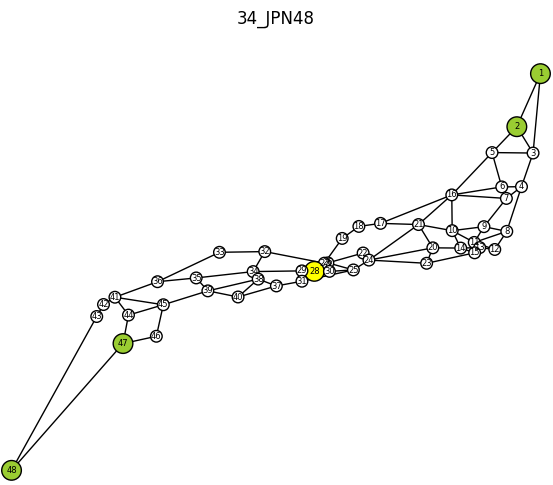}
\caption{\textit{USA100} and \textit{JPN48}, two topologies in Cluster 4.}
\label{fig:cluster4}
\end{figure}

Finally, Cluster 4 (\textit{globally favourable topologies}) consists of well-connected, large topologies
(mean connectivity 8.46\% higher, number of nodes and edges 20.13\% and 27.02\% higher than global mean), 
allowing for good performance of all protocols.
For instance, note the high connectivity and large size of \textit{USA100} and \textit{JPN48}, as shown in Figure \ref{fig:cluster4}.

The clustering results were obtained using worst-case user locations in the network (maximising the sum of pairwise shortest path distances). It would also be interesting to generalise these findings by considering more general or random locations in the network. We expect clustering to be largely insensitive to user location for Clusters 1, 3, and 4, as changing user nodes does not significantly alter the underlying graph structure on which the routing problem is solved. However, specifically for Cluster 2, the distribution rate is highly influenced by a small number of long links that limit performance. Since the results presented here place the users at the endpoints of these links, relocating the users may remove this constraint, and the protocol performance would be more similar to that of Cluster 3. We leave clustering with more general sets of user locations for future work.

\subsection{Repeater Trimming and Resource Allocation}
 
Table \ref{tab:active_reps} shows the mean percentage of active repeaters per cluster. Significant variations can be observed. For example, Cluster 1 has 81.11\% of active repeaters for MPT compared to 93.52\% of Cluster 2 and the global average of 88.86\%.

\begin{table}[H] 
\centering 
\caption{\bf Mean percentage of active nodes per cluster and protocol}
\begin{tabular}{crrrr} 
\hline
 & \multicolumn{1}{c}{SPS} & \multicolumn{1}{c}{SPT} & \multicolumn{1}{c}{MPS} & \multicolumn{1}{c}{MPT} \\
\hline
Cluster 1 & 66.67\% & 58.21\% & 83.49\% & 81.11\% \\
Cluster 2 & 71.87\% & 65.42\% & 85.70\% & 93.52\% \\
Cluster 3 & 57.23\% & 51.22\% & 89.56\% & 89.31\% \\
Cluster 4 & 54.23\% & 50.87\% & 87.30\% & 91.20\% \\
\hline
All       & 60.34\% & 54.67\% & 86.93\% & 88.86\% \\
\hline
\end{tabular} 
\label{tab:active_reps} 
\end{table}

Most topologies have redundant repeaters for both MP protocols. Hence, the initial trimming step can be interpreted as the removal of all redundant repeaters, leaving only active nodes. After this initial trimming, Monte Carlo simulations are not necessary, since the presence of redundant repeaters does not affect performance in any way. 

\begin{figure}[H]
\centering
\includegraphics[width=0.99\linewidth]{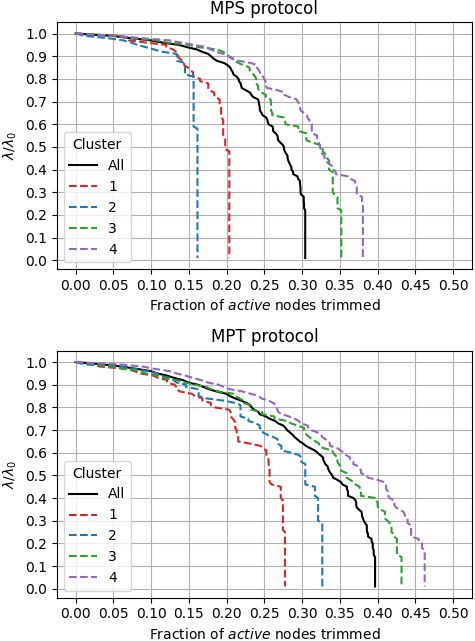}
\caption{Relationship between $\lambda/\lambda_0$ and fraction of active nodes trimmed, for MPS and MPT protocols. Colours correspond to different clusters, with black being the global average.}
\label{fig:trimmed_DRT_scaled}
\end{figure}

Figure \ref{fig:trimmed_DRT_scaled} shows the fraction $\lambda/\lambda_0$ as a function of the active nodes trimmed for MPS (top) and MPT (bottom) protocols.  Overall performance across all clusters (black line) shows that at high $\lambda/\lambda_0\geq0.85$, MPS allows for slightly more trimming of active repeaters than MPT. For instance, at $\lambda/\lambda_0=0.90$, MPT allows for trimming of only 16.21\% compared to MPS's 17.93\% (black line). MPS has repeater usage concentrated along a select few paths, and trimming repeaters outside these paths does not severely impact performance; whereas for the MPT protocol, repeater usage is distributed more evenly. However, the drop-off curve of MPS is relatively sharp, and for lower $\lambda/\lambda_0<0.85$, MPT allows for more repeater trimming than MPS. For example, at $\lambda/\lambda_0=0.50$, an average of 33.99\% of active repeaters are trimmable for MPT, compared to 27.59\% for MPS.
Once infrequently used repeaters (not part of the dominant star-based routing solutions) are trimmed, any further trimming incurs heavy performance penalties to MPS due to the disjoint routing requirement.
In contrast, the more relaxed routing constraint of MPT enables greater resilience as valid tree structures can still be formed with fewer repeaters.

For topologies in Clusters 1 and 2, trimming is relatively ineffective, 
with performance degrading rapidly as trimming progresses. This is more severe for MPS: at $\lambda/\lambda_0=0.50$, the clusters allow trimming of only 19.82\% and 16.14\% of active repeaters respectively, significantly lower than the overall average of 27.59\%. MPT also exhibits poor performance, with 25.69\% and 30.42\% of active repeaters trimmable to achieve $\lambda/\lambda_0=0.50$, moderately lower than the overall average of 33.99\%.

\begin{figure}[H]
\centering
\includegraphics[width=0.99\linewidth]{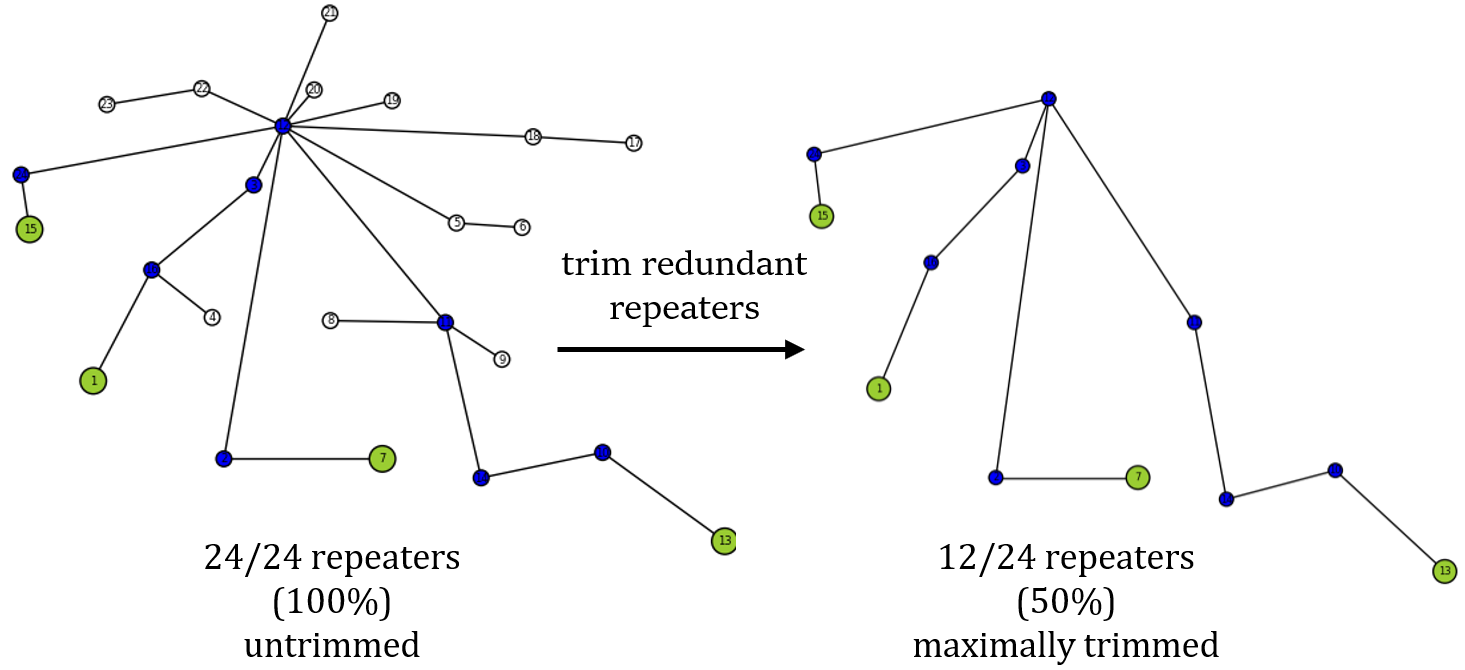}
\caption{Example repeater trimming for MPT on topology \textit{RENATER1999} (Cluster 1). Blue repeaters are used in the routing solution, whereas redundant repeaters are white. Notice the large proportion of redundant repeaters (typical of Cluster 1 topologies). Repeaters are either always or never used, hence a single trimming step results in maximal trimming. }
\label{fig:trimming_ex_RENATER}
\end{figure}

Topologies in Cluster 1 see most repeater usage concentrated on a few main routes, due to low density and connectivity. For MPS, once detours and dead ends are trimmed (around 12\% of active repeaters), arms of the star-like routing solution cannot be removed at all, so no more than 20\% of active repeaters can be trimmed. This is slightly better for MPT, but there still exist a few valid routing options; after initial trimming (around 22\% of active repeaters), subsequent trimming forces detours that incur heavy performance penalties. Figure \ref{fig:trimming_ex_RENATER} shows the trimming process in a Cluster 1 topology (\textit{RENATER1999}).

For topologies in Cluster 2, a majority of active repeaters are not trimmable, simply because topologies are small and consist of fewer nodes (for example, SPS solutions utilise 71.87\% of repeaters on average compared to global mean of 60.34\%). This is the worst-performing cluster regarding trimming for MPS, with only 16\% trimmable active repeaters; MPT exhibits slightly better (but still poor) performance due to the ability of trees to take detours around main routes.

Trimming is effective for both MP protocols on topologies in Cluster 3, but more so for MPS than MPT. At $\lambda/\lambda_0=0.50$: 32.73\% of active repeaters are trimmable, moderately higher than the global 27.59\%. For MPT, 35.92\% active repeaters are trimmable, slightly higher than the global 33.99\%.
Since most edges are of similar length, trimming of less-used active repeaters does not incur a heavy penalty, with various `detours' being of similar cost. This leads to a more gradual drop-off curve than that observed for Clusters 1 and 2. 
For MPS, heavily used repeaters are concentrated near the centre-node. Active repeaters nearer to the outskirts of the graph are rarely used, trimmable without a significant impact on performance.
On the other hand, uniform lengths imply there is no particularly dominant Steiner tree for MPT, and repeater usage is relatively spread out. This means that `less-used active repeaters' are still used a moderate amount, and trimming them non-negligibly affects performance.

Finally, trimming is extremely effective for topologies in Cluster 4 for both MP protocols (e.g.\ for $\lambda/\lambda_0=0.50$, 32.41\% and 38.43\% trimmable active repeaters compared to global 27.59\% and 33.99\% respectively). High connectivity and size mean that there are multiple routing solutions for all routing strategies. Hence, even after trimming of certain repeaters, there remain reasonable disjoint arms or Steiner trees for routing. Edge lengths are more sparsely distributed than those in Cluster 3, and thus, there are more infrequently used active repeaters, whose removal does not severely affect performance.

Overall, these results demonstrate that the impact of repeater trimming is fundamentally topology-dependent. In sparse, low-connectivity networks (Clusters 1 and 2), repeater usage is concentrated along structurally critical bottlenecks; once less-used peripheral nodes are removed, further trimming quickly disrupts essential routing structure and heavily impacts performance. In contrast, highly connected or structurally redundant networks (Clusters 3 and 4) allow substantial trimming without severe performance degradation, as multiple similar-cost routing solutions remain available.

One possible area of further work is to extend the repeater trimming protocol by identifying unnecessary swapping operations within selected paths. In networks with imperfect fidelity or swapping operations, higher performance may be achieved by bypassing intermediate repeater nodes along a path of repeaters \cite{azuma2023QuantumRepeater}. 

\section{Conclusion}
\label{sec:conclusion}

In this paper, we evaluated how network topology shapes the performance of multipartite GHZ distribution by benchmarking four routing protocols (SPS, SPT, MPS, MPT) on 81 real optical network graphs, demonstrating that protocol performance is heavily dependent on network topology.


Through $k$-means clustering of topologies using expected waiting times $\mathbb{E}[T]$ of protocols, we identified four interpretable performance classes, which we then correlated to graph-level attributes and metrics. This allows us to identify the best approach, such as enabling multi-path search or changing the routing strategy (star-based to tree-based),  depending on structural properties of networks, including the length of critical edges and the abundance of alternate paths.



We then connect performance to cost by analysing how multi-path protocols actually use repeaters, simulating iterative trimming of low-usage repeaters. Trimming tolerance is also topology-dependent. In bottlenecked or smaller graphs (Clusters 1 and 2), trimming quickly disrupts essential structure and is relatively ineffective, whereas in more uniform or well-connected graphs (Clusters 3 and 4), more substantial repeater trimming is possible without severe performance loss.

Overall, this work establishes a topology-aware perspective on multipartite routing and repeater resource allocation, providing practical guidance for topology-informed protocol selection and infrastructure planning, and ultimately enabling the design of scalable, cost-aware quantum networks in the future.


\section*{Funding}
Financial support from the EPSRC Centre for Doctoral Training (CDT) in Delivering Quantum Technologies (EP/S021582/1) is gratefully acknowledged.

\section*{Acknowledgments}
J.O.\ acknowledges support from the CDT in Delivering Quantum Technologies' Summer Placement Programme. E.S.\ completed this work while affiliated with University College London (UCL) as a student of the same CDT. E.S.'s current affiliation is Nu Quantum Ltd.
The authors thank Anuj Gore for helpful discussions and guidance in the early stages of the project.

\bibliography{biblio}

\end{document}